\documentclass[10pt,aps,twocolumn,floatfix,prl,twoside,tightenlines,nofootinbib,showpacs,superscriptaddress]{revtex4}

\usepackage[sort&compress]{natbib}
\usepackage{graphicx,amsmath,amssymb,amsopn,bm,wasysym}
\usepackage{color}

\begin{document}

\title{Low Mass Dimuons Produced in Relativistic Nuclear Collisions} 

\author{J\"org Ruppert}
\affiliation{Department of Physics, McGill University, 3600 University Street, Montreal, QC, H3A 2T8 Canada}
\author{Charles Gale}
\affiliation{Department of Physics, McGill University, 3600 University Street, Montreal, QC, H3A 2T8 Canada}
\author{Thorsten Renk}
\affiliation{Department of Physics, PO Box 35 FIN-40014 University of Jyv\"askyl\"a, Finland\\ Helsinki Institute of Physics, PO Box 64 FIN-00014, University of Helsinki, Finland}
\author{Peter Lichard}
\affiliation{Institute of Physics, Silesian University in Opava, Bezru\v{c}ovo n\'am. 13, 746 01 Opava, Czech Republic\\ Institute of Experimental and Applied Physics, Czech Technical University,
Horska 3, 12800 Prague, Czech Republic}
\author{Joseph I. Kapusta}
\affiliation{School of Physics and Astronomy, University of Minnesota, Minneapolis, MN, 55455}

\definecolor{Trot}{rgb}{1.0, 0.0, 0.0}     
\definecolor{Lblue}{rgb}{0.0,0.0,1.0}
\newcommand{\RED}[1]{\textcolor{Trot}{#1}}
\newcommand{\BLUE}[1]{\textcolor{Lblue}{#1}}

\date{\today}

\begin{abstract}
The NA60 experiment has measured low-mass muon pair production in In-In collisions at 158 A GeV with unprecedented precision.  We show that this data is reproduced very well by a dynamical model with parameters scaled from fits to measurements of hadronic transverse mass spectra and Hanbury-Brown and Twiss correlations in Pb-Pb  and Pb-Au collisions at the same energy.  The data is consistent with in-medium properties of $\rho$ and $\omega$-mesons at finite temperature and density as deduced from empirical forward-scattering amplitudes.  Inclusion of the vacuum decay of the $\rho$-meson after freeze-out is necessary for an understanding of the mass and transverse momentum spectrum of dimuons with $M \apprle 0.9~{\rm GeV}/c^2$. 
\end{abstract}


\maketitle

The main goal of the relativistic nuclear collision program is to produce and study strongly interacting matter at high temperature and density. It is hoped that exotic many-body effects may be uncovered, one of them being a quark-gluon plasma (QGP), a state where hadronic matter exhibits partonic behavior. However, other interesting phases may also manifest themselves.  In this context, electromagnetic observables  - real and virtual photons - constitute a privileged class of probes because of the near absence of final state effects. The radiation will travel essentially unscathed from its production point to the detectors. As the system expands and cools, a quantitative understanding of the net electromagnetic spectrum requires a detailed understanding of the local emissivity as well as knowledge of the space-time evolution of the radiating matter. 

The recent NA60 experiment at the CERN Super-Proton Synchrotron (SPS) has measured the production of low-mass muon pairs in In-In collisions at 158 A GeV. In this experiment the spectra of invariant mass ($M$) in the region $M \apprle 1.5$ GeV/$c^2$, and of transverse momentum ($p_T$), were obtained with unprecedented precision \cite{NA60data}.  While the invariant mass spectrum is essential to characterize in-medium changes to the electromagnetic 
current-current correlation function, the transverse momentum spectrum of dileptons is especially sensitive to the interplay between production processes and collective transverse flow. A simultaneous description of $p_T$ and $M$ spectra therefore constitutes a stringent test of the dynamical evolution of the produced matter and of our understanding of in-medium modifications of vector mesons as revealed through thermal dilepton production. In this paper we pursue a theoretical interpretation of the recent NA60 data that involve folding microscopic dilepton emission rates with a dynamical evolution model. The local pair production rate may be written 
as \cite{KGbook} 
\begin{displaymath}
\label{eqn1}
\frac{dN}{d^4x d^4q}=\frac{\alpha^2}{12 \pi^4} P(M) R(M,\vec{q}) f_{\rm B}(q_0,T)
\end{displaymath}
where $\alpha$ is the electromagnetic fine-structure constant, $T$ is temperature, and $M$, $q_0$ and $\vec{q}$ are the dimuon mass, energy and momentum, respectively. The function $P(M)$ accounts for the phase-space reduction due to the finite rest mass of the muon and does not depend on the medium's properties. The function $f_{\rm B}$ is the Bose-Einstein distribution.  The virtual photon spectral function (averaged over polarizations), $R(M,\vec{q})$, is directly related to the retarded in-medium electromagnetic current-current correlator $\Pi^{\mu \nu}_{\rm em}$ via $R=-(4\pi/M^2) {\rm Im} \Pi^\mu_{{\rm em}, \mu}$.

In the low invariant mass region, $M \apprle 1~{\rm GeV}/c^2$, the in-medium modifications of this correlator in the hadronic phase are directly related to spectral properties of light vector mesons ($\rho$, $\omega$ and $\phi$) through vector meson dominance (VMD) \cite{sakurai}. Of these, the contribution from the $\rho$-meson is the largest. Our focus here is on contributions from the $\rho$ and $\omega$ mesons; the modifications of the $\phi$-meson will be studied elsewhere. The in-medium vector meson spectral densities are evaluated in an approach where it is assumed that the vector-isovector and vector-isoscalar fields are modified mainly through scattering from nucleons and pions in the heat bath \cite{Kapusta}.  One can infer the finite temperature and density dependence of the spectral functions in the single scattering limit by extracting the leading term in the self-energy expansion.  More specifically, the self-energy is related to forward scattering amplitudes which are evaluated in a two-component model that involves the excitation of s-channel resonances upon a background - dual to the Pomeron - which becomes important especially at high energies \cite{Kapusta,Martell}. Alternatively, effective hadronic Lagrangian \cite{Rapp:1999ej,RG} or chiral reduction techniques \cite{DZ} may be used. 
\begin{figure}[t]
\begin{center}
\centerline{\includegraphics[width=0.45\textwidth]{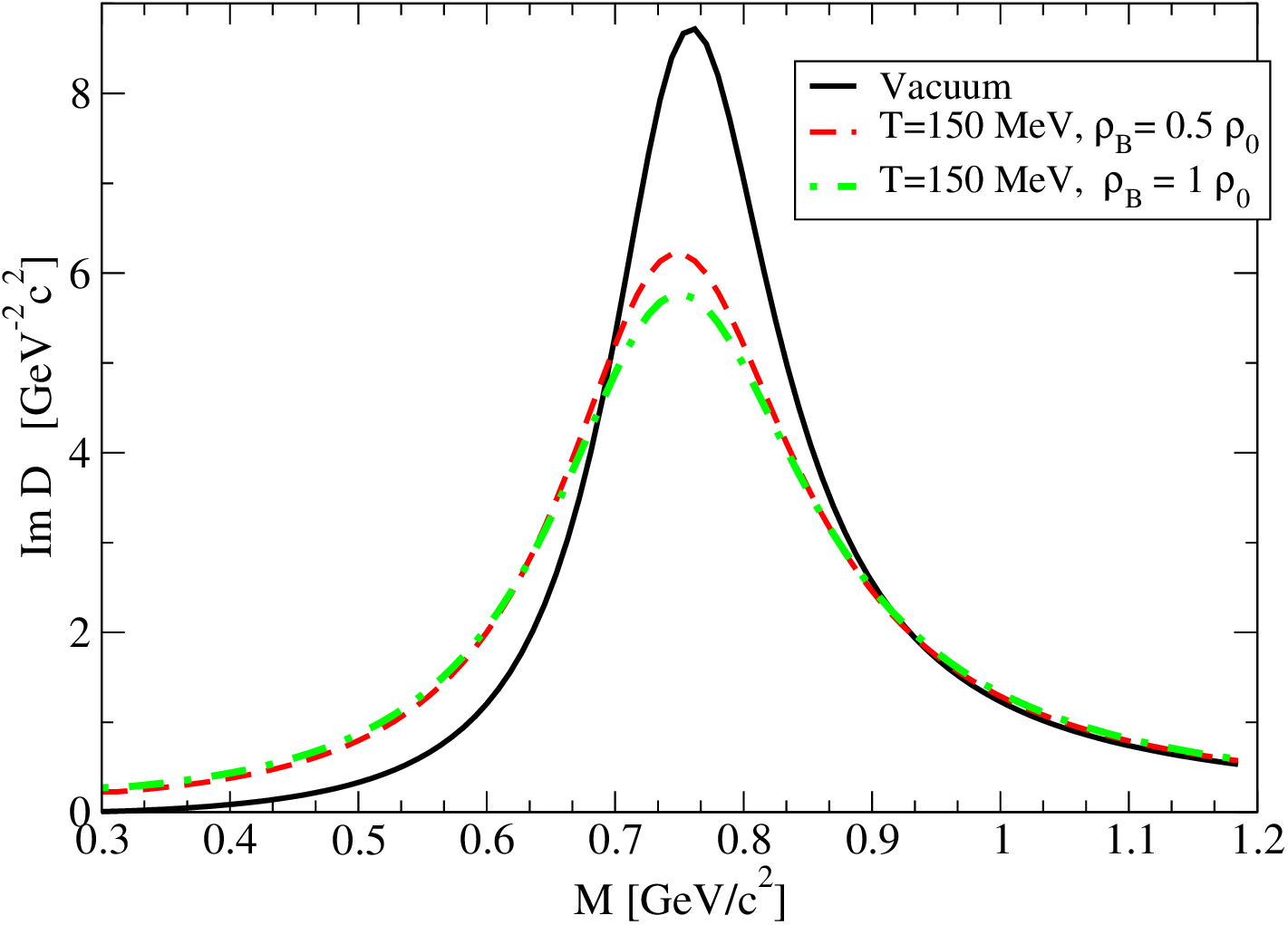}}
\end{center}
\caption{The imaginary part of the $\rho$-meson propagator for temperature $T$ = 150 MeV and baryonic density 0.5 and 1 (in units of nuclear matter density). The imaginary part of the vacuum propagator is also shown.}
\label{spectral}
\end{figure}
The results from both methods yield a negative contribution to the pole mass from interactions with pions and a positive contribution from interactions with nucleons; the net deviation from the vacuum mass being small (at most a few tens of MeV) at temperatures and densities such as those probed in 158 A GeV In-In collisions. The spectral width, however, is increased considerably owing to interactions with the medium \cite{Kapusta}; these features are shown in Figure \ref{spectral}. Collisions with nucleons are the dominant effect but collisions with pions also contribute \cite{Kapusta,Martell}.

Two additional sources are emission from a thermalized partonic phase and from four-pion annihilation processes.  A deconfined phase existing at early times and high temperatures is modeled by means of a quasiparticle model \cite{FireballMunich}.  Previous studies have made clear that quantifying dimuon emission in the invariant mass region $M \apprge 1~{\rm GeV}/c^2$ requires the inclusion of four-pion states \cite{ligale}. In the present work the required matrix elements were evaluated in the Lagrangian approach of Ref. \cite{Lichard} where the inverse process, $e^+e^-$ annihilation in the vacuum into four charged pions, was studied. The model was shown to provide reasonable agreement with the measured cross sections. In addition, we included the annihilation of two neutral and two charged pions within the same framework supplemented by intermediate states containing the $\omega$ and $h_1$ mesons.

Two contributions other than radiation from the thermalized expanding medium are essential to the understanding of the experimentally observed spectra in the mass region $M \apprle 1.5~{\rm GeV}/c^2$: the dileptons emitted from hadronic decays after the system has blown apart, and those from correlated charm decays ($D$, $\bar{D}$). The late hadronic decay contribution (for example, the spectral profile contributions from vacuum $\phi$ and $\omega$ decay) has been subtracted in the $M$ and $p_T$ distribution by NA60, but not that of the freeze-out $\rho$'s. The spectra analyzed by NA60 therefore reflect not only the {\it excess} dimuon spectrum, but also include a substantial contribution from the decay of vacuum $\rho$-mesons outside of the medium after hadronic 
freeze-out.  Since those $\rho$-mesons are strongly influenced by transverse flow, it is essential to go beyond the statistical model formulation and calculate their momentum spectrum after decoupling using the Cooper-Frye formula \cite{OurModel}. This again stresses the need for a realistic description of the spatial and temporal evolution of the system. 

A detailed description of the fireball evolution model for nucleus-nucleus collisions at the CERN SPS can be found in \cite{RenkModel,OurModel}. The main  assumption is that an equilibrated system is formed a short time $\tau_0$ after the nuclear impact. The fireball subsequently expands isentropically until mean free paths exceed the system size and particles free-stream to the detectors. This is assumed to occur at a freeze-out time (or temperature) that is the same for all hadronic species.  The entropy density $s(\tau,\eta_s,r)$ is described by the product of two Woods-Saxon distributions $s=N R(r,\tau) H(\eta_s,\tau)$ that depend on the spacetime rapidity $\eta_s=\frac{1}{2} {\rm ln} \left(\frac{t+z}{t-z}\right)$ and the transverse-plane radius $r$. The $N$ is a normalization constant. The Woods-Saxon profiles $R(r,\tau)=\left(1+{\rm exp}\left[\left(r-R_c(\tau)\right)/d_{\rm ws}\right]\right)^{-1}$ and $H(\eta_s,\tau)=\left(1+{\rm exp}\left[\left(|\eta_s|-H_c(\tau)\right)/\eta_{\rm ws}\right]\right)^{-1}$ are characterized by the thickness parameters $d_{\rm ws}$ and $\eta_{\rm ws}$ and by the size of the emitting zone as a function of proper time via $R_c(\tau)$ and $H_c(\tau)$.  The latter two functions are calculated under the assumption of constant radial and longitudinal acceleration, respectively. This translates into the rapidity of the fireball front being $\eta_s(\tau)=\eta_0+a_\eta \tau$. The parameter $2\eta_0$ is the initial size occupied by the fireball and $a_\eta$ is a longitudinal expansion parameter. The initial radial extension $R_c(0)$ is determined in a calculation of the initial density profile using the Glauber model.  Transverse flow is described best if transverse rapidity $\rho_T$ scales like $\sqrt{r}$
\cite{PTpaper1}. The accelerated longitudinal expansion driven by strong initial compression  implies that in general space-time and momentum rapidity are not the same; their mismatch $\Delta \eta$ can be seen as a characterization of how much the solution departs from the ideal Bjorken \cite{BJ} scenario. 
The parameters $\tau_0$, $a_{\perp}$,  $a_\eta$, $\eta_0$, together with the decoupling temperature $T_f$, set the scale of the spacetime evolution, and $d_{\rm ws}$ and $\eta_{\rm ws}$ specify the details of the entropy density.

The parameters listed above should in principle be fit to hadronic data and then used to predict the dimuon spectra.  However, the necessary hadronic data for In-In collisions is not yet available; fortunately, the data for Pb-Pb and Pb-Au collisions at the same beam energy are \cite{NA49,CERES}.  The same theoretical framework as described here has proven successful in the description of photon and dilepton emission and charmonium supression in those collisions \cite{RenkModel}. For the collisions of In on In at the SPS, the total entropy in semi-central In-In collisions at $\eta=3.8$ measured by NA60 \cite{NA60data} is obtained from that in peripheral (30\%) Pb-Au collisions with $2.1<\eta<2.55$ measured by CERES \cite{CERES2} by multiplying by the ratio of charged particle rapidity densities $dN_{\rm ch}/d\eta$ measured in both experiments.   The number of participant baryons and initial spatial extent are obtained via geometrical nuclear overlap calculations, while $\eta_0$ is determined under the assumption that stopping power scales approximately with the number of binary collisions per participant. The electromagnetic emission near midrapidity turns out to be quite insensitive to changes in the values of $\eta_{\rm ws}$ and $d_{\rm ws}$: these are therefore not modified.  The parameters of the accelerated expansion $a_{\perp}$ and $a_\eta$ as well as the equilibration time are assumed to be primarily determined by the incident system energy so they are kept as in Pb-Pb collisions.  The largest uncertainty is the choice of the decoupling temperature. Because the In-In system is smaller than Pb-Pb, a higher decoupling temperature is expected. However, to give an unambiguous answer would only be possible with simultaneous measurements of HBT correlations and transverse mass spectra. Here we choose $T_f=130$ MeV. As alluded to earlier, kinetic equilibrium of all processes is assumed until this universal decoupling temperature is reached. The fact that the four-pion processes contribute all the way down to $T_f$ translates into an upper limit to their contribution, as microscopic descriptions of the dynamics (see, for example, \cite{Nonaka}) suggest a sequential decoupling of the different channels.  The equation of state in the hadronic phase and the chemical potentials $\mu_{\pi}, \mu_{K}$ are inferred from statistical model calculations as described in \cite{RenkModel,Renk:2002sz}.  The resulting fireball has a peak temperature of about 250 MeV and a lifetime of about 7.5 fm/$c$. 

The thermal contributions have to be calculated by folding the space-time evolution of the expanding matter with the thermal rates. 
For the mass spectra this amounts to
\begin{equation}
\nonumber
\frac{dN}{M dM d\eta}= \int d^4x \int d\psi \int dp_T p_T {\cal A}(M,p_T,\eta)  \frac{dN}{d^4x  d^4q}
\end{equation}
where ${\cal A}$ represents the detector acceptance of NA60. The space-time evolution enters via the dependence of the thermal rates on temperature, baryon and pion chemical potential, and energy and momentum of the decaying virtual photon in the local rest frame. The equilibrium thermal rates have been augmented by appropiate fugacity factors ${\rm exp}(n \mu_{\pi}/T)$ with $n=2,3,4$ for the thermal $\rho$, $\omega$, and four-pion annihilation contributions. 

\begin{figure}[t]
\begin{center}
\vspace{0.2cm}
\centerline{\includegraphics[width=0.50\textwidth]{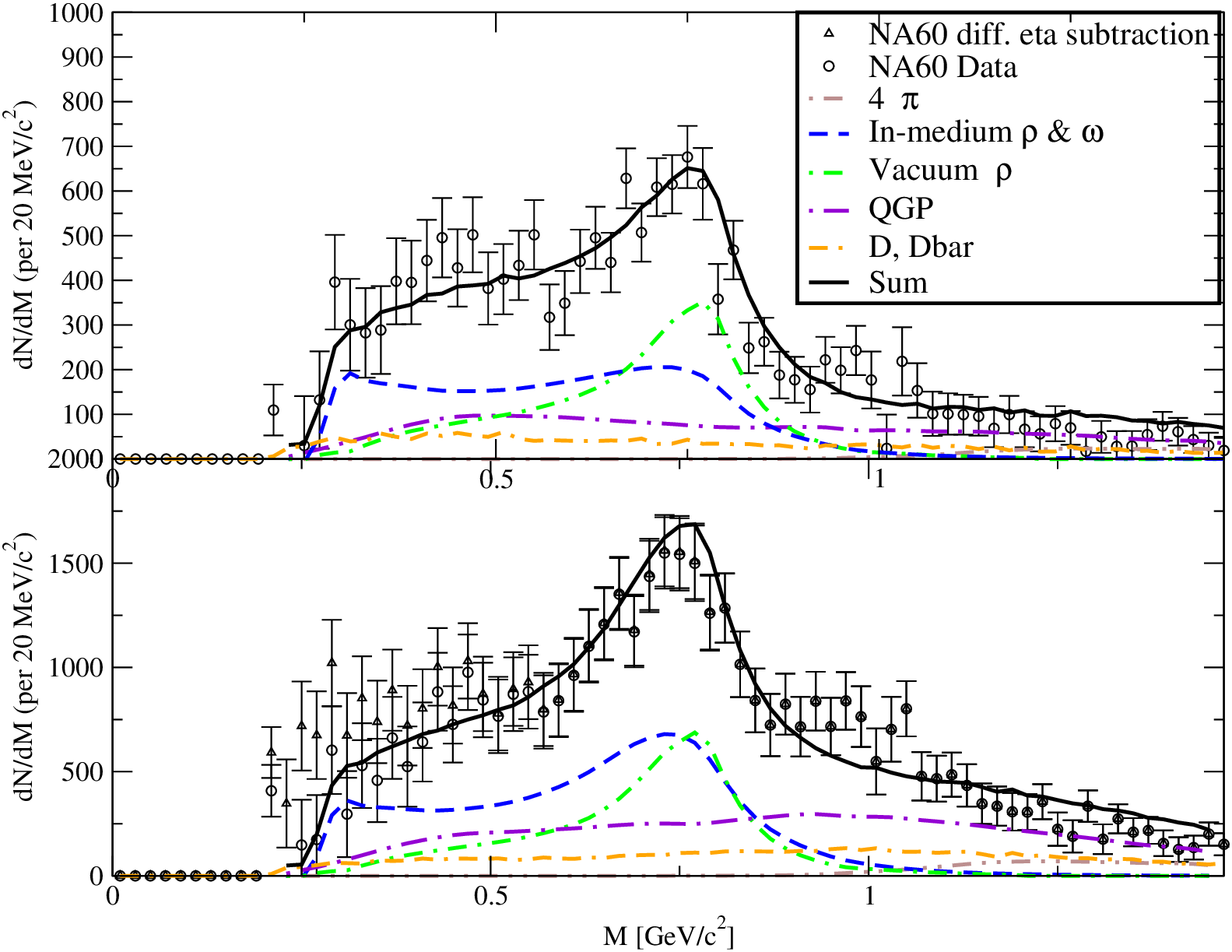}}
\end{center}
\caption{Mass spectrum in semi-central In-In collisions at SPS \cite{NA60data} compared to theory. The lower panel shows an integration over all $p_T$, the upper panel one over high transverse momenta $1~{\rm GeV}/c<p_T<2~{\rm GeV}/c$. Partial contributions arise from $\rho$ decays in vacuum after freeze-out, thermal in-medium $\rho$ and $\omega$ decays, radiation from a thermalized QGP and from thermal four-pion annihiliation, and from correlated open charm decay.}
\label{mspectra}
\end{figure}

A comparison of a theoretical calculation of the invariant mass spectra, with its different components, with NA60 measurements in semi-central In-In collisions at the SPS is presented in Fig. \ref{mspectra}.
Consider the all $p_T$-data in the lower panel first. 
The region below $1~{\rm GeV}$ and at masses smaller than the vacuum mass of the $\rho$ meson clearly demonstrates a considerable in-medium broadening of the $\rho$ and $\omega$ mesons. The decays of the vacuum $\rho$ after freeze-out are important, and in certain mass ranges are of the same order as the in-medium $\rho$ and $\omega$ meson decay contributions.  The contribution from the QGP phase becomes more important and eventually dominant with higher $M$. We find that four-pion annihilation processes are subdominant even for masses above $M \apprge 1.25~{\rm GeV}/c^2$.
This differs from the findings in \cite{vanHeesRapp} where the four-pion annihilation was derived in the soft pion limit, assuming chiral mixing.  
 
Even though the spectrum is integrated over all $p_T$, and mass is Lorentz invariant, a successful theoretical understanding of these spectra still requires a detailed understanding of the $p_T$ spectra since the acceptance ${\cal A}$ is transverse momentum-dependent.  Additional information is obtained if one compares the theoretical prediction for momentum cuts of the mass spectrum with the experimental data.  Since the acceptance restricts considerably the information that can be obtained from the  mass spectrum at low transverse momenta, this contribution is not shown even though our model does provide a good description of the spectrum in this window of $0<p_T<0.5~{\rm GeV}/c$.  At higher transverse momenta, $1~{\rm GeV}/c<p_T< 2~{\rm GeV}/c$, the relative contribution of dimuons from vacuum $\rho$-decays is relatively enhanced as the emission then occurs at a later stage of the evolution where considerable transverse flow has already built up. 
\begin{figure}[t]
\begin{center}
\centerline{\includegraphics[width=0.50\textwidth]{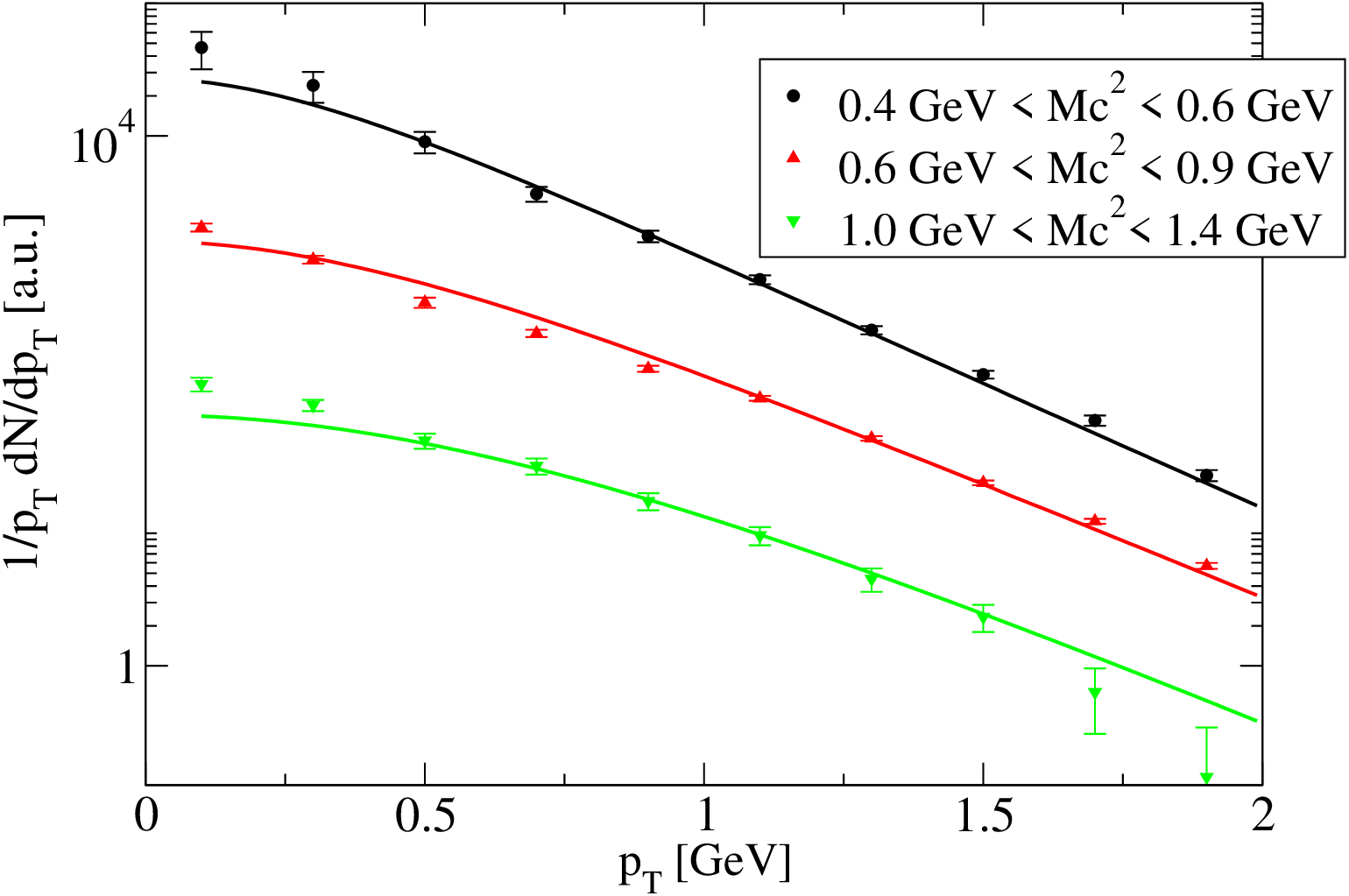}}
\end{center}
\caption{Transverse momentum spectra as obtained from theory for semi-central In-In collision. The acceptance-corrected data are from NA60 \cite{NA60data}.}
\label{pspectra}
\end{figure}
The robustness of our theoretical understanding of the collision can be further assessed from a comparison with the acceptance corrected $p_T$ spectra for different mass windows: see also \cite{PTpaper1}. Figure \ref{pspectra} shows the $p_T$ spectra in three different mass windows for semi-central collisions. (The data are averaged over different centrality classes excluding peripheral collisions \cite{NA60data}.  We also performed this averaging in \cite{PTpaper1,PTpaper2} and found that differences between the averaged data and the semi-central collision data are small.) 

The mass region $0.4~{\rm GeV}/c^2<M<0.6~{\rm GeV}/c^2$ and the $\rho$-like mass region $0.6~{\rm GeV}/c^2<M<0.9~{\rm GeV}/c^2$ receive most of their contribution from the late hadronic stages, namely, from decays of in-medium vector mesons and the vacuum $\rho$ mesons after freeze-out. In those stages considerable flow has already been built up which implies that the blueshift of the spectra by flow is large.  The difference between the $\rho$-like and the lower mass region in the spectra is caused by the different contributions of the in-medium vector mesons and the vacuum $\rho$. The latter receives the maximum flow and predominantly contributes to the $\rho$-like region for momenta above $\sim 1~{\rm GeV}/c$.  The slope of the transverse momentum spectrum in the mass region $1.0~{\rm GeV}/c^2<M<1.4~{\rm GeV}/c^2$ is dominated by contributions from the early QGP phase where flow has not yet built up while a contribution from four-pion annihilation processes is subdominant.  If four-pion annihilation processes were more substantial this would result in considerable hardening of the $p_T$-spectrum which is not observed \cite{NA60data,PTpaper1,PTpaper2}.

In conclusion, we have shown that low mass dimuon production as measured in 158 A GeV In-In collisions at the CERN SPS reflects substantial in-medium broadening of the $\rho$ meson spectral function in the hot and dense nuclear medium. Furthermore, we found that at higher invariant masses thermal radiation with $T>170~{\rm MeV}$ dominates over four pion annihiliation processes. This is especially relevant to a theoretical understanding of intermediate mass dimuon production.

\textit{Acknowledgments.} We are grateful to S. Damjanovic for help concerning the implementation of the NA60 acceptance and to her and H. Specht for discussions. This work was supported by the Natural Sciences and Engineering Research Council of Canada, by the Academy of Finland, by Czech Ministry of Education grant LC07050, and by DOE grant DE-FG02-87ER40328.

\vspace{-0.5cm}

\begin{flushleft}

\end{flushleft}


\end{document}